\documentstyle[prb,preprint,
aps,psfig]{revtex}
\begin{document}
\draft


\title{Structural and electronic properties of an azamacrocycle, 
$C_{26}H_{18}N_{6}$}
\author{K. Doll$^*$ and G. Zwicknagl}
\address{Institut f\"ur Mathematische Physik, TU Braunschweig,
Mendelssohnstra{\ss}e 3, D-38106 Braunschweig}
\address{$^*$Electronic address: k.doll@tu-bs.de}

\maketitle

\begin{abstract}
We compute the structure of an azamacrocycle, $C_{26}H_{18}N_{6}$. 
Two approximatively planar elliptical structures with
$C_2$ or $C_i$ symmetry are found to be nearly degenerate.
The roughly circular conformation observed in metal complexes turns out 
to be $\sim$ 0.6 eV higher in energy. We suggest that this difference
is mainly due to electrostatic interactions. 
We discuss the results on various levels of theory
(Hartree-Fock, local density and gradient corrected 
density functional calculations).

\end{abstract}

\pacs{ }

\narrowtext
\section{Introduction}

In recent years, the formation of metal complexes of macrocyclic ligands
has become an important research field which is rapidly developing.
The interest in this field originates from the possibility to build
supramolecular architectures and thus to finally arrive at tailor-made
materials with novel properties. The goal to design new materials with
a combination of desired properties, however, requires a quantitative
understanding of the properties of the individual units, their response
to chemical substitutions as well as their interactions. Of prime
importance is to understand the molecular structure and the factors which
determine it. 
The azamacrocycle, $C_{26}H_{18}N_6$,
has been widely used as a ligand, for the formation of complexes
with a wide variety of metals such as
alkali metals\cite{Bell1984,Bell1991}, lanthanides\cite{Benetollo1991,Aruna1996}, 
earth alkaline metals\cite{Drew1979} , lead\cite{Drew1979}, cadmium
\cite{Drew1979,Marchetti1989} or yttrium\cite{Aruna1996}. 
In all these metal complexes, the azamacrocycle is
approximatively ring-like, with the metal ion bonding to the six nitrogen
atoms.
It was surprising, when the free ligand was synthesized
and was found to be approximatively planar-elliptical
\cite{Bell1986}. 

A full explanation of the difference of the two structures has not yet
been given. However, modern codes and computers make ab-initio 
calculations
on the Hartree-Fock (HF) and
density-functional level for these systems possible. We 
therefore investigated the different structures and their electronic
properties. The main target of this article is a quantitative explanation of
the relative stabilities of these structures. Thus we optimized
the geometry and computed
the total energy, charge density  and  the vibrational spectrum.
In addition, we try to assess the overall validity of the methods and the
approximations involved. We compare the results for Hartree-Fock and
various density functional calculations to estimate the ambiguities introduced
by the simplified description of electronic correlations. In the following
 section (\ref{methodsection}) we give
details about the method. The results are presented
in section \ref{resultssection}, and finally we summarize the article
in section \ref{summarysection}.

\section{Method}
\label{methodsection}

All the calculations were done with the code  Gaussian 98\cite{Gaussian98}.
We employ the Hartree-Fock method,
the local density approximation (LDA) (with the correlation
functional V from reference \onlinecite{VWN}) and the gradient corrected
Hybrid functional B3LYP \cite{B3LYP}. 
The bulk of the calculations was done with a 6-31G Gaussian 
basis set for C, H and N atoms. Additional
calculations with an enhanced 6-31G$^{**}$ basis set (i.e.
with an additional $p$-polarization 
function for H and an additional $d$-polarization function
for C and N) were performed to investigate the dependence of the results
on the choice of the basis set. 
 As an optimization of the system in its
 crystal structure (monoclinic, space group P2$_1$/C) is presently
prohibitive, we performed a calculation on the isolated
molecule. 
The effects of this approximation will be discussed in section 
\ref{resultssection}.
As a first step, a full geometry optimization with respect to
all parameters and without symmetry was performed. 
Depending on the geometry at the beginning of the optimization process, 
the final optimized geometry turned
out to be either $C_2$, $C_i$ or $C_{2v}$ symmetric, the three minima
considered in this article.
To ensure the stability
of the optimized 
structures, in addition vibrational frequencies were calculated.

\section{Results}
\label{resultssection}

\begin{figure}
\caption{Optimized structures, side view}
\label{vonseiteps} 
\centerline{\psfig{figure=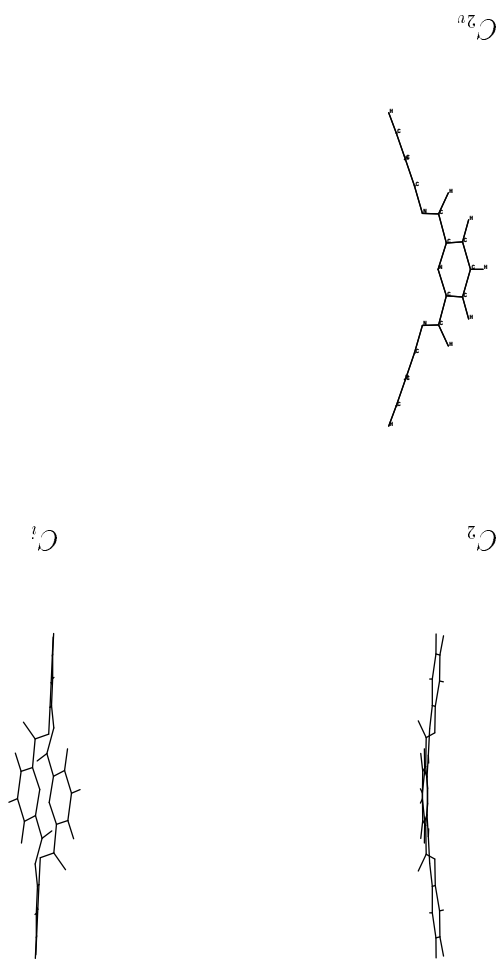,width=16cm,angle=0}}
\end{figure}

The results of the calculations are summarized in tables
\ref{Energytable},\ref{Geometrieexp}, \ref{GeometrieCi}
and \ref{Geometriesaddle}.
First, we discuss the optimized geometries.
The first calculated structure is visualized in figure \ref{elliptischC2ps}
using Molden \cite{Molden}.
This structure is approximatively
planar-elliptical and has $C_2$ symmetry. Thus, it
is the one which has the same symmetry as a single $C_{26}H_{18}N_6$ unit of
the experimentally observed structure \cite{Bell1986} 
(i.e. the inner
hydrogens, 41 and 42 in figure \ref{elliptischC2ps}
are on the same face of the molecule).
The second structure
(structure 2) 
is also approximatively planar-elliptical, but has $C_i$ symmetry.
In the view in figures \ref{elliptischC2ps}, it looks virtually identical
as structure 1, and 
the difference compared to structure 1 can better be seen in figure
\ref{vonseiteps}.
The alternative to these flat structures is the approximatively
circular structure (structure 3) which is found to be another local minimum. 
We display the latter structure in figure
\ref{Sattelvonobenps}.
This latter structure is saddle-like; in a view from the top
it appears approximatively circular (figure \ref{Sattelvonobenps}),
and, when viewing from the side, we note
that there is a significant bending (figure \ref{vonseiteps}).

The energy differences of the structures on the HF,  LDA and B3LYP level
of theory are displayed in table \ref{Energytable}. 
We find that the saddle-like
 structure 3 is highest in energy at all levels of theory. While HF
gives the highest energy difference, this is reduced by LDA and even
more by B3LYP. The calculation with the best basis set used in this
article (6-31 G$^{**}$) gives an energy difference of 0.6 eV at the B3LYP
level. Including zero-point vibrations hardly changes the energy difference,
as the zero-point energy was found to be 0.384 $E_h$
(1 $E_h$= 27.2114 eV)
for  structure 1 and 2 and 0.383 $E_h$ for the saddle-like 
structure 3
(B3LYP, 6-31G$^{**}$ basis set). The computed vibrational
frequencies were in the range
from $\sim$20 to $\sim$ 3200 cm$^{-1}$. The energy difference is also
consistent with earlier results based on molecular mechanics
calculations \cite{Bell1991}, where, depending on the approximations,
an energy difference in the region from $\sim$ 0 to 0.7 eV was obtained.
Comparing the results for structure 1 and 2, we note that at the HF
level, structure 1 is found to be slightly lower than structure 2.
This changes when the calculations are performed at the density
functional level: now,
structure 2 is consistently lower than structure 1. This energy
 splitting ranges from +0.0003 $E_h$ to -0.002 $E_h$, 
The quality of the basis set has a minor influence,
and zero-point vibrations are negligible for this splitting.
 
Although B3LYP calculations are usually reliable,
this contradiction of the various levels of theory
makes the correct prediction of the molecular
conformation very difficult. Concerning the molecular crystal, lattice
effects may well be of the order of the magnitude of the splitting
between structure 1 and structure 2. A safer prediction of the
structure of the molecular crystal would thus require an optimization of the 
periodic system which is presently prohibitive.

Structural parameters, together with Mulliken populations are displayed
in tables \ref{Geometrieexp}, \ref{GeometrieCi} and \ref{Geometriesaddle}.
The charge distributions indicate that all the hydrogen atoms carry
charges of $\sim$ 0.1 $|e|$, and the nitrogen atoms
 $\sim -0.5 |e|$. The carbon atoms 
carry charges from $\sim$ -0.1 to $\sim 0.25|e|$;
obviously they are stronger negative for those atoms with a hydrogen neighbor
and positive for those atoms with a nitrogen neighbor.
The charge distribution is nearly consistent for all the structures. 

We further note that the structural parameters vary only weakly
when using basis sets of different quality. 
Similarly, the charges varied slightly
by about 0.1 $|e|$.  This deviation may be viewed as
an error bar and the conclusions are not affected.

A comparison with the experimental structure data was made in table
\ref{relativeerror}. We used the standard deviation of the distance
matrix 

$
\sqrt{\frac{2}{n(n-1)}\sum_{i=1}^n \sum_{j=1}^{i-1}
\frac{(d_{i,j}^{computed}-d_{i,j}^{exp})^2}{(d_{i,j}^{exp})^2}}
$

as a measure for the error of the computed geometry (with $n$ the
number all atoms and $d_{i,j}^{computed}$ and $d_{i,j}^{exp}$ being
the computed and experimental distances between the atoms $i$ and $j$,
respectively). We note that the standard deviation is very small both for
the structure with $C_2$ symmetry as well as the structure with 
$C_i$ symmetry. However, the structure with the $C_2$ symmetry  found
experimentally has  a slightly higher standard deviation.
Finally, the structural parameters of structure 3
 with $C_{2v}$ symmetry deviate strongly.

Comparing the  structures, we can thus conclude that electrostatic
effects favor the approximatively planar  structures 
1 and 2 over structure 3: the main effect seems
to be that the two inner hydrogen atoms (41  and 42 in figure 
\ref{elliptischC2ps}) have lowered
the energy of the molecule by inwards bending and pointing towards the
negatively charge nitrogen atoms. This additional attraction would, for
example, for the hydrogen 42 and nitrogen 6, be
of the order 0.17 $\times$ 0.45 /4 a.u. $\sim$ 0.5 eV and thus seems to be 
the likely explanation for the preference of  structure 1. 
In addition, we note that the hydrogen atoms (33, 50  and 34, 49) in structure
1 would feel a relatively strong repulsion in a strictly planar structure,
and a slight distortion of the molecule thus lowers the energy. This is
indeed found in the experiment and in our simulations. A similar 
distortion is also found in structure 2.
Finally, in structure 3 the hydrogen atoms (39, 43; 41, 45; 40, 44 and 
42,46) would feel
a repulsion in a planar structure; and this repulsion can be reduced
by a distortion to a saddle-like structure.

In summary, we feel that the preference of the planar elliptical
 structures over the saddle-like
 can be explained by a purely electrostatic argument. Embedding
a positively charged metal ion was experimentally
found to change the situation: the
saddle-like structure is now the preferred one. 
We feel that this effect may have a similar reason:
in the approximatively planar elliptical structures 1 or 2, the 
additional repulsion from the two inwards bending hydrogen atoms
would make it difficult to embed a cation, and moreover, the saddle-like
structure offers more space to embed a cation. In the saddle-like 
structure, the metal ion feels the electrostatic attraction due to all
six nitrogen atoms. However, a full quantitative analysis 
of a system with metal ions will be more difficult because the individual
units are charged and thus also 
effects from the anions in the molecular crystal must be included.

\section{Summary}
\label{summarysection}
We have demonstrated that  ab-initio  Hartree-Fock and density-functional 
calculations can explain the relative stabilities of flat versus
saddle-like
conformations of the azamacrocycle $C_{26}H_{18}N_6$. The charge distribution
indicates that electrostatic
effects are the reason for the  energy difference
of $\sim$ 0.6 eV between these conformations. 
However, two flat structures with $C_2$ symmetry and $C_i$ symmetry were 
found to be nearly degenerate.
The distance matrices of
the structures with $C_2$ symmetry and
with $C_i$ symmetry are both in very good agreement with the distance matrix
from  experimental data. Finally, we showed
that the energetically highest structure has a geometry similar to the
one observed in systems with a metal embedded in the azamacrocycle.
We feel that this saddle-like structure is preferable for the embedding
of a metal ion because of the charge distribution and the available space.

The main limitation of these calculations was the restriction to 
molecules instead of molecular crystals. New code developments
(for example the CRYSTAL code with analytical gradients\cite{IJQC})
will hopefully make calculations on the molecular crystal feasible,
as well as more demanding simulations on crystals with embedded metal
ions\cite{Bell1984,Bell1991,Benetollo1991,Aruna1996,Drew1979,Marchetti1989}.

\section{Acknowledgments}
The authors would like to thank Prof. S. Laschat and Dr. C. Th\"one for
helpful discussions.

\onecolumn

\newpage
\begin{table}
\begin{center}
\caption{\label{Energytable}
Relative energies, computed with respect to the energy of the structure with
$C_2$ symmetry, at different levels of theory (1 $E_h$=27.2114 eV)}
\vspace{5mm}
\begin{tabular}{ccccc}
method & basis set & \multicolumn{2}{c}{relative energy with respect to}  & \\
 & & \multicolumn{2}{c}{the structure with $C_2$ symmetry $[E_h]$}\\
& & structure with & structure with \\
& &     $C_i$ symmetry & $C_{2v}$ symmetry\\
HF & 6-31G &  0.0003 & 0.043 \\
HF & 6-31G$^{**}$ & 0.0009 & 0.026   \\
LDA & 6-31G &  -0.0016 & 0.038 \\
LDA & 6-31G$^{**}$ &  -0.0021 & 0.028 \\
B3LYP & 6-31G &  -0.0008 & 0.033 \\
B3LYP, with zero-point vibrations & 6-31G & -0.0008 & 0.031 \\
B3LYP & 6-31G$^{**}$ & -0.0016 & 0.022 \\
B3LYP, with zero-point vibrations & 6-31G$^{**}$  & -0.0015 & 0.022 \\
\end{tabular}
\end{center}
\end{table}

\newpage
\begin{table}
\begin{center}
\caption{\label{Geometrieexp}
Optimized geometrical parameters and Mulliken populations, 
B3LYP, 6-31G$^{**}$ basis set, approximatively
elliptical structure (1) with C2 symmetry
(i.e. simultaneously 
x $\rightarrow$ -x and y$\rightarrow$ -y is a symmetry operation.
Only positions of symmetry unique atoms are given.}
\vspace{5mm}
\begin{tabular}{cccccc}
atom number & type & x & y & z & charge \\
 & & [\AA] &  [\AA] & [\AA] & $|e|$ \\
   1,4 & N  &     0.352 &  -2.027 &  -0.144 & -0.52 \\
   2,5 & N  &    -3.187 &  -1.795 &  -0.022 & -0.53 \\
   3,6 & N  &    -2.939 &   1.147 &   0.275 & -0.45 \\
   7,8 & C  &     1.376 &  -2.890 &  -0.236 &  0.26 \\
  9,10 & C  &     1.184 &  -4.269 &  -0.393 & -0.11 \\
 11,12 & C  &    -0.122 &  -4.765 &  -0.407 & -0.06 \\ 
 13,14 & C  &    -1.184 &  -3.877 &  -0.295 & -0.09 \\
 15,16 & C  &    -0.900 &  -2.504 &  -0.181 &  0.24 \\
 17,18 & C  &    -1.948 &  -1.465 &  -0.114 &  0.12 \\
 19,20 & C  &    -4.252 &  -0.889 &   0.112 &  0.24 \\
 21,22 & C  &    -5.522 &  -1.494 &   0.104 & -0.10 \\
 23,24 & C  &    -6.692 &  -0.765 &   0.288 & -0.09 \\
 25,26 & C  &    -6.619 &   0.613 &   0.502 & -0.10 \\
 27,28 & C  &    -5.380 &   1.247 &   0.498 & -0.11 \\ 
 29,30 & C  &    -4.190 &   0.530 &   0.278 &  0.27 \\
 31,32 & C  &    -2.738 &   2.313 &  -0.206 &  0.09 \\
 33,34 & H  &    -5.315 &   2.312 &   0.696 &  0.09 \\
 35,36 & H  &    -7.521 &   1.191 &   0.680 &  0.09 \\
 37,38 & H  &    -7.653 &  -1.271 &   0.288 &  0.09 \\
 39,40 & H  &    -5.549 &  -2.569 &  -0.035 &  0.10 \\
 41,42 & H  &    -1.569 &  -0.443 &  -0.128 &  0.18 \\
 43,44 & H  &    -2.218 &  -4.201 &  -0.307 &  0.12 \\
 45,46 & H  &    -0.302 &  -5.831 &  -0.511 &  0.10 \\
 47,48 & H  &     2.038 &  -4.935 &  -0.490 &  0.10 \\ 
 49,50 & H  &     3.539 &  -2.923 &  -0.650 &  0.09 \\
\end{tabular}
\end{center}
\end{table}

\newpage
\begin{table}
\begin{center}
\caption{\label{GeometrieCi}
Optimized geometrical parameters and Mulliken populations, 
B3LYP, 6-31G$^{**}$ basis set, approximatively
elliptical structure (1) with $C_i$ symmetry
(i.e. simultaneously 
x $\rightarrow$ -x, y$\rightarrow$ -y and $z\rightarrow -z$
is a symmetry operation.
Only positions of symmetry unique atoms are given.}
\vspace{5mm}
\begin{tabular}{cccccc}
atom number & type & x & y & z & charge \\
 & & [\AA] &  [\AA] & [\AA] & $|e|$ \\
 1,4 & N  &          0.309 &   -1.984 & -0.414 &  -0.52 \\
 2,5 & N  &         -3.220 &   -1.804 & -0.122 &  -0.52 \\
 3,6 & N  &         -2.926 &    1.035 &  0.547 &  -0.45 \\
 7,8 & C   &         1.343 &   -2.838 & -0.326 &   0.26 \\
 9,10 & C  &         1.166 &   -4.225 & -0.214 &  -0.11 \\
11,12 & C  &        -0.132 &   -4.738 & -0.200 &  -0.06 \\
13,14 & C  &        -1.204 &   -3.857 & -0.261 &  -0.08 \\
15,16 & C  &        -0.934 &   -2.479 & -0.360 &   0.23 \\
17,18 & C  &        -2.005 &   -1.461 & -0.358 &   0.15 \\
19,20 & C  &        -4.283 &   -0.914 &  0.055 &   0.24 \\
21,22 & C  &        -5.564 &   -1.483 & -0.073 &  -0.10 \\
23,24 & C  &        -6.721 &   -0.752 &  0.162 &  -0.09 \\
25,26 & C  &        -6.625 &    0.580 &  0.576 &  -0.10 \\
27,28 & C  &        -5.375 &    1.172 &  0.712 &  -0.10 \\
29,30 & C  &        -4.192 &    0.464 &  0.425 &   0.26 \\
31,32 & C  &        -2.702 &    2.263 &  0.269 &   0.09 \\
33,34 & H  &        -5.295 &    2.194 &  1.070 &   0.09 \\
35,36 & H  &        -7.520 &    1.151 &  0.805 &   0.09 \\
37,38 & H  &        -7.693 &   -1.224 &  0.056 &   0.09 \\
39,40 & H  &        -5.611 &   -2.531 & -0.350 &   0.09 \\
41,42 & H  &        -1.657 &   -0.439 & -0.508 &   0.15 \\
43,44 & H  &        -2.235 &   -4.189 & -0.220 &   0.12 \\
45,46 & H  &        -0.299 &   -5.808 & -0.125 &   0.10 \\
47,48 & H  &         2.028 &   -4.882 & -0.139 &   0.09 \\
49,50 & H  &         3.490 &   -2.942 &  0.088 &   0.09 \\
\end{tabular}
\end{center}
\end{table}

\newpage
\begin{table}
\begin{center}
\caption{\label{Geometriesaddle}
Optimized geometrical parameters and Mulliken populations,
B3LYP, 6-31G$^{**}$ basis set, C$_{2v}$ symmetric (i.e. x
$\rightarrow$ -x and y $\rightarrow$ -y are symmetry operators). Only 
positions of symmetry unique atoms are given.}
\vspace{5mm}
\begin{tabular}{cccccc}
atom number & type & x & y & z & charge \\
   1,2,3,4    &    C     &    -1.150  &  3.105 &   0.695 &  0.26 \\
   5,6,7,8    &    C     &    -1.198  &  4.323 &   1.398 & -0.11 \\
   9,10       &    C     &     0.000  &  4.942 &   1.744 & -0.06 \\
  11,12       &    N     &     0.000  &  2.518 &   0.343 & -0.47 \\
  13,14,15,16 &    C     &     2.416  &  2.412 &   0.359 &  0.09 \\
  17,18,19,20 &    N     &     2.436  &  1.347 &  -0.342 & -0.39 \\
  21,22,23,24 &    C     &     3.628  &  0.711 &  -0.685 &  0.22 \\
  25,26,27,28 &    C     &     4.789  &  1.395 &  -1.079 & -0.10 \\
  29,30,31,32 &    C     &     5.930  &  0.698 &  -1.473 & -0.09 \\
  33,34,35,36 &    H     &     2.153  &  4.764 &   1.665 &  0.09 \\
  37,38       &    H     &     0.000  &  5.885 &   2.283 &  0.10 \\
  39,40,41,42 &    H     &     3.329  &  2.863 &   0.780 &  0.07 \\
  43,44,45,46 &    H     &     4.771  &  2.480 &  -1.116 &  0.09 \\
  47,48,49,50 &    H     &     6.810  &  1.244 &  -1.799 &  0.08 \\
\end{tabular}
\end{center}
\end{table}

\newpage
\newpage
\begin{table}
\begin{center}
\caption{\label{relativeerror}
Error of the distance matrix of the three
conformations with respect to the experimental geometry.}
\vspace{5mm}
\begin{tabular}{cccc}
structure & method & basis set   &  standard deviation \\
1 ($C_2$ symmetry) &
HF & 6-31G &   0.025 \\
1 & HF & 6-31G$^{**}$ & 0.029 \\
1 & LDA & 6-31G &  0.029 \\
1 & LDA & 6-31G$^{**}$ & 0.026 \\
1 & B3LYP & 6-31G &  0.028 \\
1 & B3LYP & 6-31G$^{**}$ & 0.024  \\
2 ($C_i$) &
HF & 6-31G$^{**}$ & 0.019  \\
2 & LDA & 6-31G$^{**}$ & 0.026 \\
2 & B3LYP & 6-31G$^{**}$ & 0.022  \\
3 ($C_{2v}$) &
B3LYP & 6-31G$^{**}$ & 0.15 \\
\end{tabular} 
\end{center}  
\end{table}   

\newpage
\begin{figure}
\caption{Optimized planar-elliptical 
structure with $C_2$ symmetry (structure 1). Structure 2 with $C_i$ symmetry
looks virtually identical in this view.}
\label{elliptischC2ps} 
\end{figure}
\centerline{\psfig{figure=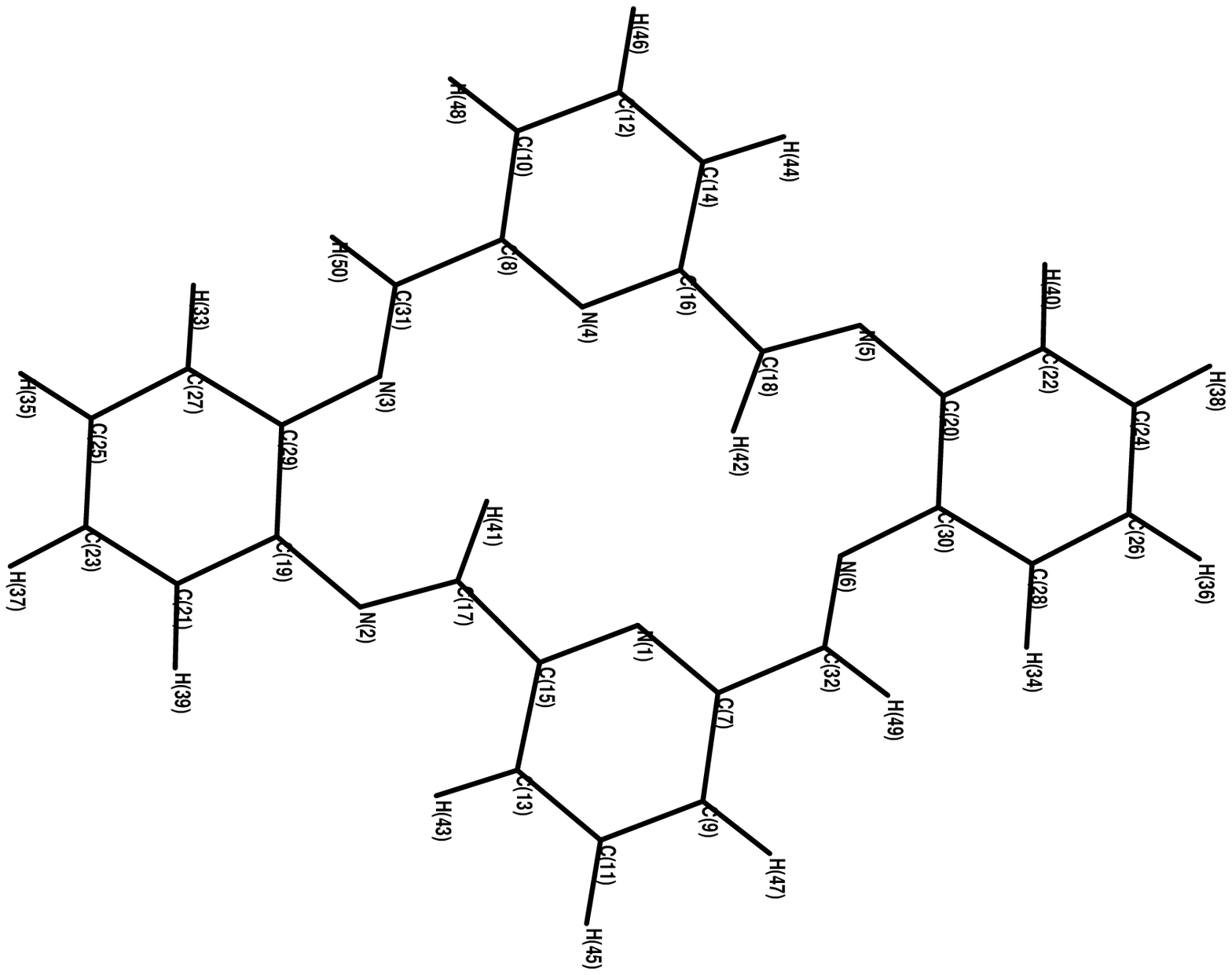,width=18cm,angle=90}}
\vfill

\newpage
\begin{figure}
\caption{Optimized saddle-like structure with $C_{2v}$ symmetry
(structure 3)}
\label{Sattelvonobenps} 
\end{figure}
\centerline{\psfig{figure=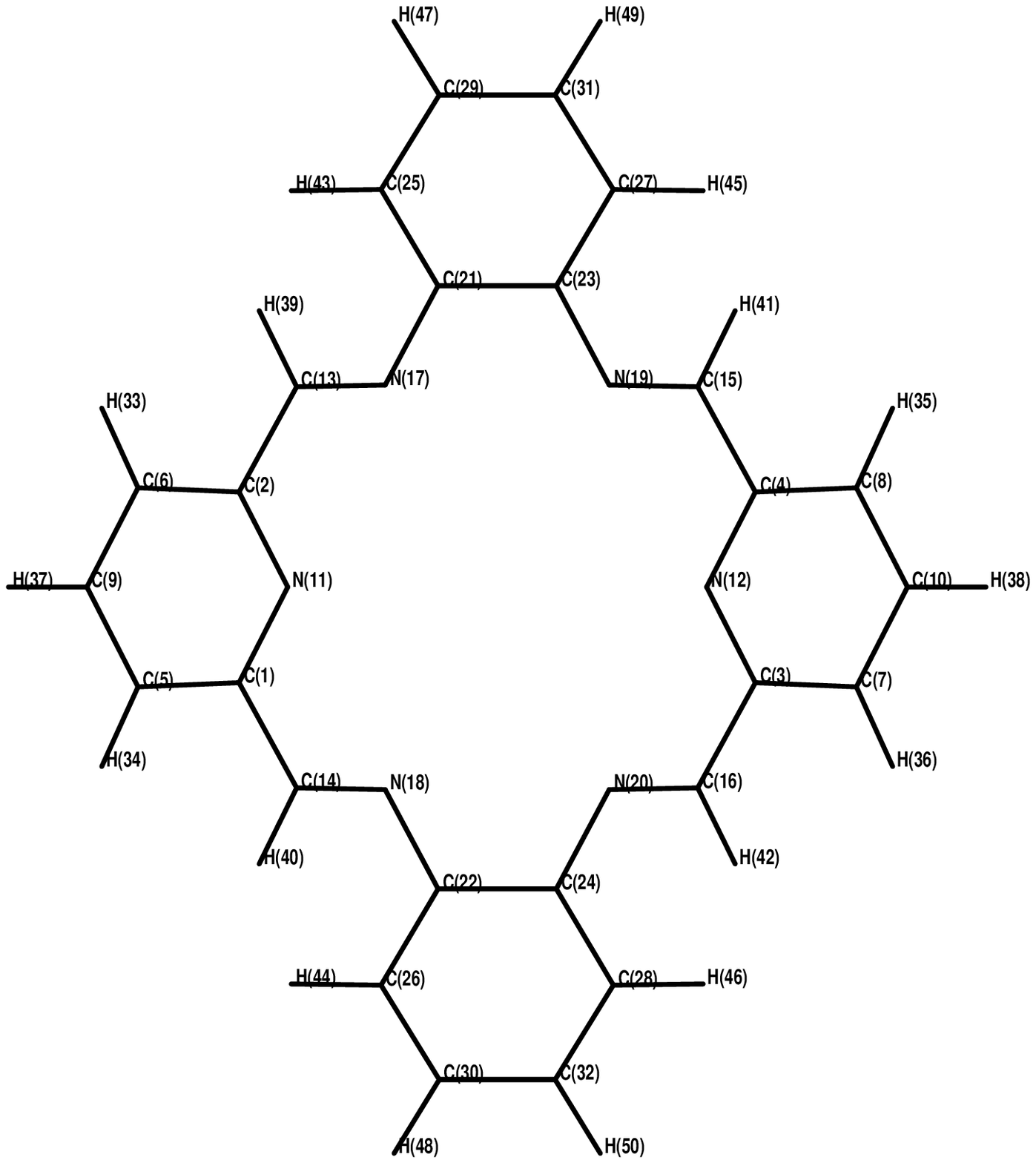,width=18cm,angle=90}}
\vfill

\newpage
\vfill

\end{document}